\documentclass[letterpaper, 10 pt, conference]{ieeeconf}  % Comment this line out if you need a4paper

\IEEEoverridecommandlockouts                              % This command is only needed if 
\usepackage{graphicx}
\usepackage{url}  %IVM added this because of the URLs in the reference list
\usepackage{pifont}

\usepackage{amsmath}
\DeclareMathOperator*{\argmax}{argmax}

\usepackage{lipsum}
\title{\LARGE \bf
Deepfake Audio Detection Using Spectrogram-based Feature \\ and Ensemble of Deep Learning Models 
}

\author{Lam~Pham$^{1*}$, 
        Phat~Lam$^{2*}$,
        Truong~Nguyen$^{3}$,
        Huyen~Nguyen$^{4}$,
        Alexander~Schindler$^{5}$       
\thanks{L. Pham and A. Schindler are with Austrian Institute of Technology, Vienna, Austria.}%
\thanks{P. Lam and T. Nguyen are with HCM University of Technology,  Ho Chi Minh city, Vietnam}%
\thanks{H. Nguyen is with Tokyo University  of Agriculture and Technology,  Tokyo, Japan}%
\thanks{(*) Main and equal contribution into the paper.}
}

\begin{document}

\maketitle
\thispagestyle{empty}
\pagestyle{empty}

%%%%%%%%%%%%%%%%%%%%%%%%%%%%%%%%%%%%%%%%%%%%%%%%%%%%%%%%%%%%%%%%%%%%
\begin{abstract}

In this paper, we propose a deep learning based system for the task of deepfake audio detection.
In particular, the draw input audio is first transformed into various spectrograms using three transformation methods of Short-time Fourier Transform (STFT), Constant-Q Transform (CQT), Wavelet Transform (WT) combined with different auditory-based filters of Mel, Gammatone, linear filters (LF), and discrete cosine transform (DCT).
Given the spectrograms, we evaluate a wide range of classification models based on three deep learning approaches.
The first approach is to train directly the spectrograms using our proposed baseline models of CNN-based model (CNN-baseline), RNN-based model (RNN-baseline), C-RNN model (C-RNN baseline).
Meanwhile, the second approach is transfer learning from computer vision models such as ResNet-18, MobileNet-V3, EfficientNet-B0, DenseNet-121, SuffleNet-V2, Swint, Convnext-Tiny, GoogLeNet, MNASsnet, RegNet.
In the third approach, we leverage the state-of-the-art audio pre-trained models of Whisper, Seamless, Speechbrain, and Pyannote to extract audio embeddings from the input spectrograms.
Then, the audio embeddings are explored by a Multilayer perceptron (MLP) model to detect the fake or real audio samples.
Finally, high-performance deep learning models from these approaches are fused to achieve the best performance.
We evaluated our proposed models on ASVspoof 2019 benchmark dataset.
Our best ensemble model achieved an Equal Error Rate (EER) of 0.03, which is highly competitive to top-performing systems in the ASVspoofing 2019 challenge.
Experimental results also highlight the potential of selective spectrograms and deep learning approaches to enhance the task of audio deepfake detection.

\indent \textit{Items}--- deepfake audio, deep learning model, spectrogram, ASVspoof dataset.
\end{abstract}
%%%%%%%%%%%%%%%%%%%%%%%%%%%%%%%%%%%%%%%%%%%%%%%%%%%%%%%%%%%%%%%%%%%%%%%%%%%%%%%%
\section{INTRODUCTION}
\label{intro}

Sound-based applications represent a revolutionary paradigm in the rapidly evolving landscape of Internet of Sound (IoS) technology, where audio signals serve as the primary medium for data transmission, control, and interaction among interconnected devices \cite{ios_01, ios_02}. 
Voice-activated module in an IoS system, such as smart home devices, voice banking, home automation systems, and virtual assistants, relies on recognizing the user's voice to activate critical functions and generally involve confidential information. 
However, with the advancement of deep learning technologies, the emergence of spoofing speech attacks, commonly referred to as 'Deepfake', has become more prevalent. 
These attacks involve various AI-based speech synthesis techniques (e.g., Speech to Text \cite{intro_s2t}, Voice Conversion \cite{intro_s2t}, Scene Fake \cite{scenefake}, Emotion Fake \cite{emotion_fake}), posing significant threats to the integrity and authenticity of voice-activated systems. 
Consequently, the detection of audio deepfakes has become a crucial area of research, drawing considerable attention from the research community. 
Several benchmark datasets and following challenges such as ASVspoof \cite{top3}, Audio Deep synthesis Detection (ADD) \cite{ADD}, have been proposed, which facilitates the creation of various systems and techniques to handle this task. 
Existing studies can be divided into two kind: pipeline solutions (consisting of a front-end feature extractor and a back-end classifier) and end-to-end solutions \cite{survey1}.
The top-performing systems using these two methods in the ASVspoof and ADD competitions are mainly score-level fusion systems~\cite{survey1}. 
However, these systems lack a comprehensive evaluation of how individual spectrograms and classifiers affect overall performance, which is crucial for further research motivation and research direction. 
Other successful systems utilize deep features through various supervised embedding methods, such as DNNs \cite{deep_feature1} and RNNs \cite{deep_feature_2}. 
Despite their effectiveness, these embeddings are trained on specific datasets and may encounter the issues of overfitting and susceptibility to adversarial attacks. 
This reduces the model's ability to generalize to new, unseen data, particularly when the dataset is not sufficiently large or diverse. 
Meanwhile, other approaches that can manage generalization and domain adaptation, such as transfer learning and leveraging embeddings from large pre-trained audio models, have not been extensively explored. 
To tackle these mentioned limitations, we therefore propose an ensemble of deep learning based models for audio deepfake detection task, which is achieved via a comprehensive analysis in terms of multiple spectrogram-based features and deep learning approaches. Our key contributions can be highlighted as: 
\begin{itemize}
    \item Evaluated the efficacy different spectrograms in combination with auditory filters to model performance.    
    \item Evaluated a wide range of architectures leveraging both transfer learning and end-to-end networks.
    \item Explored the performance of audio embeddings extracted from state-of-the-art pre-trained models (e.g. Whisper, Speechbrain, Pyannote) on deepfake detection.
    \item Proposed an ensemble model via selective spectrograms and models from experiment, indicating the research focuses for further improving the task of deepfake audio detection. 
\end{itemize}

\section{Proposed Deep Learning Based Systems}
\label{systems}

\begin{figure}[t]
    \centering
    \includegraphics[width =1.0\linewidth]{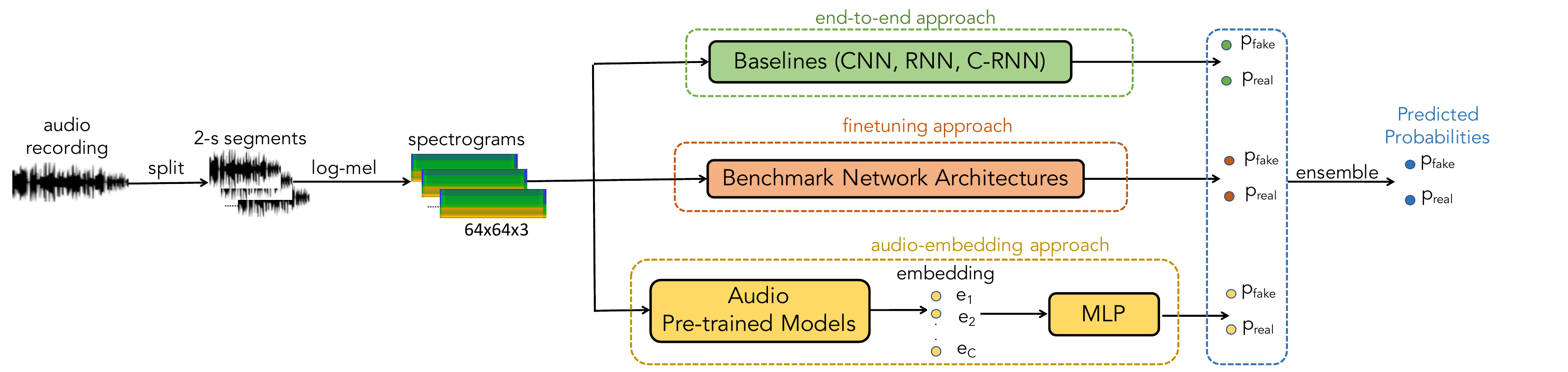}
           	\vspace{-0.6cm}
	\caption{The high-level architecture of proposed deep learning based system for deepfake audio detection}
       	\vspace{-0.4cm}
    \label{f1}
\end{figure}

The high-level architecture of proposed deep learning based system for audio deepfake detection, which is denoted in Fig.~\ref{f1}, comprises two main parts: front-end spectrogram-based feature extraction and back-end deep learning model for classification.
In particular, the draw input audio recordings are first split into 2-second segments. This segment length generally provides sufficient context to capture important features and allows faster training and inference for applications requiring real-time detection. 
%For this reason, we train our proposed systems on 2-second audio segments. 
Next, the 2-second audio segments are transformed into spectrograms.
Finally, the spectrograms are explored by back-end deep learning models to detect real or fake audio segments.

There are three deep learning based approaches are proposed in this paper.
The first approach is shown in the upper part in Fig.~\ref{f1}.
In this approach, referred to as the end-to-end approach, proposed models are used to train input spectrograms directly.
%, then report the real or fake probabilities.
%The models which are used in the first approach are CNN-based model, RNN-based model, C-RNN-based model (i.e. these models are referred to as CNN baseline, RNN baseline, and C-RNN baseline respectively) and a wide range of network architectures of Resnet18, MobileNetV3, EfficientNetB0, DenseNet121, SuffleNetV2, Swint, ConvnextTiny, GoogleNet, Mnasnet.
In the second approach as shown in the middle part in Fig.~\ref{f1}, referred to as the finetuning approach , we fine-tune benchmark network architectures which are popularly used in the computer vision domain. 
Regarding the third approach as shown in the lower part in Fig.~\ref{f1}, we leverage the state-of-the-art pre-trained models which were trained on large audio datasets in advance.
Then, we feed spectrograms input into these audio pre-trained models to obtain audio embeddings.
The audio embeddings are finally classified into either real or fake class by a Multilayer Perceptron (MLP). 
We refer this approach to as the audio-embedding approach.
Finally, individual and high-performance models from three approaches are selected and fused to achieve the best performance.

\subsection{Spectrogram-based Feature Extraction}
\label{spec}

\begin{figure}[t]
    \centering
    \includegraphics[width =0.8\linewidth]{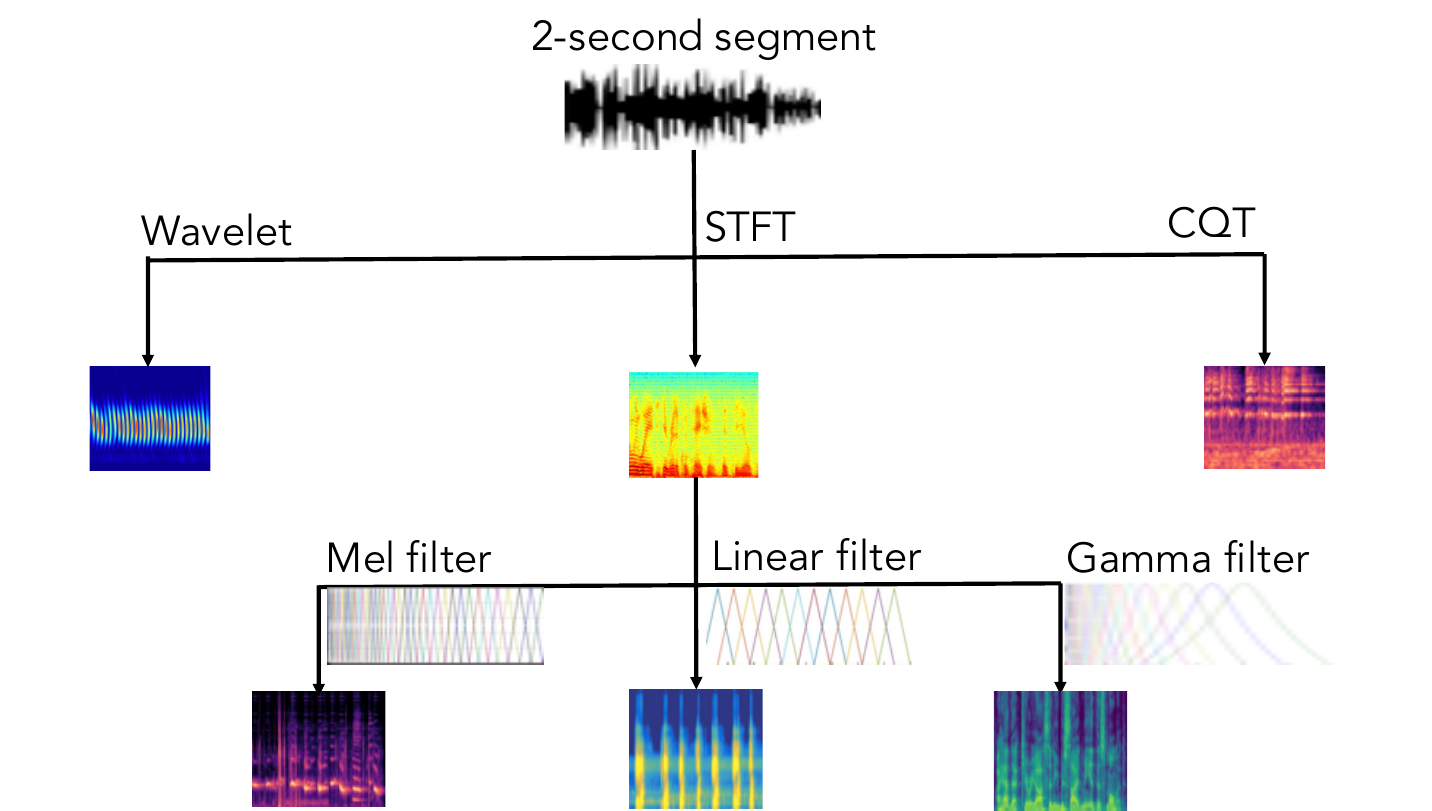}
           	\vspace{-0.3cm}
	\caption{Generate spectrograms using different spectrogram transformation methods and auditory filter models}
       	\vspace{-0.2cm}
    \label{f2}
\end{figure}

Fig.~\ref{f2} presents how 6 different spectrograms are generated in this paper.
In particular, 6 spectrograms are generated from three transformation methods of Short-time Fourier Transform (STFT), Constant-Q Transform (CQT), Wavelet Transform (WT). 
Presumably, each type of spectrogram focus on different perspectives on frequency content and might catch different inconsistencies in the audio signal. 
The combination of these spectrograms allows model to learn a broader range of features and patterns, potentially improving its ability to generalize and detect deepfakes.
Additionally, we also establish different auditory-based filters: Mel, Gammatone focus on subtle variations relevant to human auditory perception; linear filters (LF) isolates specific frequency bands.,
%and discrete cosine transform (DCT) provides a compact representation of features. 
Integrating these filters alongside pre-defined spectrograms enriches the available features and further enhances the robustness to variations of the detection system.

As we use the same settings of the window length, the hop length, the filter number with 1024, 512, 64 for all spectrograms, generated spectrograms present the same tensor shape of 64$\times$64.
Then, DCT is applied on spectrograms across the temporal dimension. 
Finally, we apply delta and delta-delta to these spectrograms, generate three dimensional tensor of 64$\times$64$\times$3 (i.e. the original
spectrogram, delta, and delta-delta are concatenated across the third dimension).

\begin{table}[t]
\caption{The CNN, RNN, and C-RNN baseline network architectures} 
       	\vspace{-0.2cm}
    \centering
    {
\scalebox{0.9}{

\begin{tabular}{|c|c|}
%\textit{Custom MobileFaceNet based network for classification} \\
\hline
\textbf{Models} & \textbf{Configuration} \\  
\hline
CNN baseline                  & \textbf{3} $\times$ \{Conv(32/64/128)-ReLU-AP-Dropout(0.2)\}   \\ 
                              & \textbf{1} $\times$ \{Dense(256)-ReLU-Dropout(0.2)\} \\
                              & \textbf{1} $\times$ \{Dense(2)-Softmax\}  \\
                              \hline
RNN baseline                  &\textbf{2} $\times$ \{BiLSTM(128/64)-ReLU-Dropout(0.2)\}   \\ 
                              & \textbf{1} $\times$ \{Dense(256)-ReLU-Dropout(0.2)\} \\
                              & \textbf{1} $\times$ \{Dense(2)-Softmax\}   \\
                              \hline
C-RNN baseline                &\textbf{3} $\times$ \{Conv(32/64/128)-ReLU-AP-Dropout(0.2)\}   \\ 
                              &\textbf{2} $\times$ \{BiLSTM(128/64)-ReLU-Dropout(0.2)\}   \\
                              &\textbf{1} $\times$ \{Dense(256)-ReLU-Dropout(0.2)\}\\
                              &\textbf{1} $\times$ \{Dense(2)-Softmax\} \\
                              \hline
\end{tabular}
}
    }
\vspace{-0.4cm}
\label{t1}
\end{table}
\subsection{End-to-end deep learning approach}
\label{model01}
Regarding the end-to-end deep learning approach, we propose three baseline models of CNN-based model, RNN-based model, C-RNN-based model, which are referred to as the CNN baseline, RNN baseline, and C-RNN baseline, respectively.
The detailed configuration of these baselines are presented in Table~\ref{t1}. CNNs are the most common architecture for this task, which can effectively capture and learn spectral features within local  frequency bands such as harmonic structures, formants, pitch variations, high-frequency artifacts, etc. Meanwhile, RNNs focus on detecting natural sequential patterns that can be disrupted in synthetic audio \cite{rnn_deepfake} (e.g. temporal coherence, prosodic features such as rhythm, stress, and intonation). Consequently, the usage of C-RNN baseline is based on the expectation of combine both spectral features and temporal features for distinguishing characteristics of deepfake audio.

\subsection{Transfer learning approach}
\label{model02}
Additionally, we also evaluate a wide range of benchmark network architectures in the computer vision domain such as ResNet-18, MobileNet-V3, EfficientNet-B0, DenseNet-121, SuffleNet-V2, Swint, Convnext-Tiny, GoogLeNet, MNASsnet, RegNet.
In particular, these networks were trained on the ImageNet1K dataset~\cite{imagenet_ds} in advance. Their pre-trained weights can capture rich and generalized features about pattern recognition in images, 
which can be potentially adapted to identifying patterns in  spectrograms via parameter finetuning. In this approach, the final dense layer of these mentioned networks is modified to match the binary classification task of deepfake audio detection before conducting the fine-tune process.

\subsection{Audio-embedding deep learning approach}
\label{model03}
In the audio-embedding deep learning approach, we leverage the state-of-the-art audio pre-trained models of Whisper~\cite{whisper}, Seamless~\cite{seamless}, Speechbrain~\cite{speechbrain}, and Pyanote~\cite{pyanote1, pyanote2}. 
These pre-trained models are utilized for their ability to capture robust and high-level feature representations of genuine speakers in practice such as pitch, tone, accent, and intonation from their diverse training data. This capability is crucial for distinguishing between real and fake audio. 
Therefore, the spectrogram inputs are first fed into these pre-trained models to obtain audio embeddings
Given the audio embeddings, we propose a Multilayer perceptron (MLP), as shown in Table~\ref{t2}, to detect real or fake audio.
%The feature map layer in the pre-trained models which present the audio embedding and the dimension of audio embeddings are presented in Table~\ref{t2}.

\begin{table}[t]
\caption{The audio pre-trained models and the Multilayer Perceptron} 
       	\vspace{-0.2cm}
    \centering
    {
\scalebox{0.8}{
\begin{tabular}{|l|c|c|}
%\textit{Custom MobileFaceNet based network for classification} \\
\hline
\textbf{Models} & \textbf{Using License} & \textbf{Embedding size/configuration} \\  
\hline
Whisper~\cite{whisper} & MIT &512 \\
SpeechBrain~\cite{speechbrain} & Apache2-0 &192 \\
SeamLess~\cite{seamless} & MIT & 1024\\
Pyannote~\cite{pyanote1, pyanote2} & MIT & 512\\
\hline
MLP      & Our proposal    &\textbf{1} $\times$ \{Dense(128)-ReLU \} \\
         &                     &\textbf{1} $\times$ \{Dense(2)-Softmax \} \\
\hline
\end{tabular}
}
    }
   	%\vspace{-0.5cm}
\label{t2}
\vspace{-0.3cm}
\end{table}
\begin{table*}[t]
    \caption{Performance comparison among deep learning models and ensemble of high-performance models \\ on Logic Access evaluation subset in ASVspoofing 2019} 
        	\vspace{-0.2cm}
    \centering
    \scalebox{0.85}{
    \begin{tabular}{|l  |c|c|c|c|c|c|} 
        \hline 
        \textbf{Systems} &\textbf{Spectrograms}   &  \textbf{Models}  &  \textbf{Acc} &  \textbf{F1} &\textbf{AuC}  &\textbf{ERR}\\  
        \hline   
         A1 & STFT       &CNN   &\textbf{0.87} &\textbf{0.89} &\textbf{0.96} &\textbf{0.08}\\
         A2 & CQT       &CNN   &0.89 &0.90 &0.92 &0.14\\
         A3 & WT        &CNN   &0.84 &0.86 &0.89 &0.17\\
         A4 & STFT \& LF    &CNN   &\textbf{0.88} &\textbf{0.90} &\textbf{0.96} &\textbf{0.08}\\
         A5 & STFT \& MEL   &CNN   &0.86 &0.88 &0.95 &0.11\\
         A6 & STFT \& GAM   &CNN   &\textbf{0.85} &\textbf{0.87} &\textbf{0.96} &\textbf{0.08}\\         
         \hline
         \hline                 
         B1 & STFT \& LF   &RNN   &0.92 &0.91 &0.88 &0.17\\
         B2 & STFT \& LF   &CRNN  &0.88 &0.90 &0.96 &0.14\\
         \hline
         \hline
         C1 & STFT \& LF   &ResNet-18   &0.49 &0.58 &0.51 &0.47\\
         C2 & STFT \& LF   &MobileNet-V3    &0.59 &0.67 &0.52 &0.48\\
         C3 & STFT \& LF   &EfficientNet-B0     &0.52 &0.61 &0.51 &0.48\\
         C4 & STFT \& LF   &DenseNet-121     &0.58 &0.66 &0.51 &0.48\\
         C5 & STFT \& LF   &ShuffleNet-V2  &0.64 &0.71 &0.53 &0.48\\       
         C6 & STFT \& LF   &Swin\_T     &\textbf{0.84} &\textbf{0.87} &\textbf{0.94} &\textbf{0.09}\\         
         C7 & STFT \& LF   &ConvNeXt-Tiny     &\textbf{0.88} &\textbf{0.90} &\textbf{0.96} &\textbf{0.075}\\         
         C8 & STFT \& LF   &GoogLeNet     &0.53 &0.62 &0.51 &0.47\\                                    
         C9 & STFT \& LF   &MNASNet    &0.62 &0.70 &0.54 &0.47\\   
         C10 & STFT \& LF  &RegNet    &0.50 &0.60 &0.50 &0.48\\   
         \hline
         \hline
         D1 & STFT \& LF   &Whisper+MLP   &\textbf{0.85} &\textbf{0.88} &\textbf{0.95} &\textbf{0.10}\\
         D2 & STFT \& LF   &Speechbrain+MLP   &0.77 &0.81 &0.81 &0.25\\
         D3 & STFT \& LF   &Seamless+MLP   &0.86 &0.88 &0.87 &0.20\\
         D4 & STFT \& LF   &Pyannote+MLP   &0.64 &0.71 &0.78 &0.27\\
       \hline
       \hline             
       A1 + A2 &STFT, CQT  &CNN  &\textbf{0.91} &\textbf{0.92} &\textbf{0.98} &\textbf{0.06}\\ 
       A1 + A3 &STFT, WT   &CNN  &0.88 &0.90 &0.96 &0.09\\ 
       A1 + A2 + A3 & STFT, CQT, WT   &CNN  &0.90 &0.92 &0.98 &0.07\\ 
       \hline                           
       A4 + A5 &LFCC, MEL &CNN  &0.88 &0.90 &0.97 &0.08\\ 
       A4 + A6 &LFCC, GAM &CNN  &0.87 &0.89 &\textbf{0.98} &\textbf{0.065}\\ 
       A4 + A5 + A6 &LFCC, MEL, GAM &CNN  &0.88 &0.90 &0.98 &0.069\\ 
       \hline
       A4 + C6 & LFCC & CNN, Swint\_T &0.87 &0.89 &0.96 &0.078 \\                                   
       A4 + C7 & LFCC & CNN, ConvNeXt-Tiny &0.88 &0.90 &\textbf{0.97} &\textbf{0.07} \\   
       A4 + C6 + C7 & LFCC & CNN, ConvNeXt-Tiny, Swint\_T &0.88 &0.89 &0.97 &0.072 \\                                                                 
       \hline
       \hline
       \textbf{A2 + A4 + A6 + C7} &\textbf{CQT, LFCC, GAM} &\textbf{CNN, ConvNeXt-Tiny, Whisper}  &\textbf{0.90} &\textbf{0.91} &\textbf{0.994} &\textbf{0.03}\\ 

       \hline 
      
    \end{tabular}
    }
    \vspace{-0.3cm}
    \label{table:R1} 
\end{table*}
\subsection{Ensemble of models}
\label{ensemble}
As an individual model works on 2-second audio segment, the predicted probability of an entire audio recording is computed by averaging of predicted probabilities over all 2-second segments.
Consider $\boldsymbol{p}^{(n)} = [p_{1}^{(n)}, p_{2}^{(n)},..., p_{C}^{(n)}]$,  with $C$ being the category number of the n-th out of \(N\) 2-second segments in one audio recording. The probability of an entire audio recording is calculated by the average classification probability which denoted as $\bar{\boldsymbol{p}} = [\bar{p}_{1},\bar{p}_{2},..., \bar{p}_{C}]$ where:
\begin{equation}
    \label{eq:mean_stratergy_patch}
    \bar{p}_{c} = \frac{1}{N}\sum_{n=1}^{N}p_{c}^{(n)}  ~~~  for  ~~ 1 \leq c \leq C 
\end{equation}

To ensemble of results from individual models, we propose a MEAN fusion.
In particular, we first conduct experiments on the individual models, then obtain the predicted probability as  \(\mathbf{\hat{p}}_{s}= (\bar{p}_{s_1}, \bar{p}_{s_2}, ..., \bar{p}_{s_C})\) where $C$ is the category number and the s-th out of \(S\) individual models evaluated. 
Next, the predicted probability after MEAN fusion \(\hat{\boldsymbol{p}}_{f-mean} = (\hat{p}_{1}, \hat{p}_{2}, ..., \hat{p}_{C}) \) is obtained by:
\begin{equation}
    \label{eq:mix_up_x1}
     \hat{p_{c}} = \frac{1}{S} \sum_{s=1}^{S} \hat{p}_{s_c} ~~~  for  ~~ 1 \leq c \leq C 
\end{equation}
Finally, the predicted label \(\hat{y}\) for an entire audio sample is determined as:

\begin{equation}
    \label{eq:label_determine}
    \hat{y} = \argmax (\hat{p}_{1}, \hat{p}_{2}, ...,\hat{p}_{C} )
\end{equation}

\section{Experiments and Results}
\label{exper}

\subsection{Datasets and Evaluation Metrics}
We evaluate the proposed models on the Logic Access dataset of ASVspoofing 2019 challenge. 
The Logic Access dataset comprises three subsets(fake sample/real sample) of `Train'(22800/2580), `Develop'(22296/2548), and `Evaluation'(63882/7355), in which fake audio were generated from 19 AI-based generative systems.
The models are trained on `Train' subset, then evaluated and saved on `Develop' subset.
Finally, the models are test on the `Evaluation' subset and the final result on this subset is reported.

%\subsection{Evaluation Metrics}
We obey the ASVspoofing 2019 challenge, then use the Equal Error Rate (ERR) as the main metric for evaluating proposed models.
We also report the Accuracy, F1 score and AuC score to compare the performance among proposed models.

%\subsection{Experimental settings}
%Our proposed deep neural networks were constructed with Pytorch framework.
%The proposed deep neural networks were trained for 60 epochs using Titan RTX 24GB GPU. 
%The Adam method~\cite{Adam} is used for the optimization and the learning rate is set to 0.001, 0.0001, and 0.0001 for the end-to-end, fine-tune, and audio-embedding deep learning approaches, respectively.

\subsection{Results and Discussion}

\textbf{Evaluation of spectrogram inputs}: Consider the efficacy of feature extraction among proposed spectrogram inputs (i.e. systems from A1 to A6), STFT outperforms other compared spectrograms (models such as A1, A4, A5 achieves the best ERR score of 0.08 while the combination of STFT \& LF obtains slightly better accuracy and F1 score of 0.88 and 0.9 respectively). This result suggests that STFT is often better suited for identify deepfake artifacts due to its uniform resolution in time and frequency \cite{stft} while the interpretable features extracted from linearly filtered signals are suitable for classification algorithms.

\textbf{Multiple deep learning approaches:} Regarding end-to-end deep learning approach (A1 to B2),  both RNN and C-RNN approaches obtains ERR score of 0.14 and 0.17, 
significantly worse than using only CNN with the best score of 0.08. 
This indicates the specific patterns indicative of deepfake audio might not be primarily temporal but rather spatial in the spectrogram representation. 
In the finetuning and audio embeddding-based approaches (C1 to C10 and D1 to D4), Swint, Convnext-Tiny and Whisper stand out as best systems within the corresponding approaches with competitive EER score of 0.09, 0.0075 and 0.10 respectively. This suggests the potential of these approaches when choosing the appropriate networks for enhancement.  

\textbf{Ensembles:} The experimental results presented in Table \ref{table:R1} underscore the significant effectiveness of ensemble techniques in detecting audio deepfakes. 
Specifically, the combination of STFT and LF spectrograms (A1+A2) achieves a score of 0.06, marking an improvement of 0.02 compared to best systems utilizing single spectrograms. 
Similarly, ensembles of models show slight enhancements such as the combination of CNN and ConvNeXt-Tiny which helps to reduce the ERR by 0.01 and 0.005 compared to individual models. 
These findings suggest that diverse feature extraction via ensembling multiple spectrograms substantially enhances overall performance compared to evaluating a wide range of models on a single spectrogram. 
Importantly, the ensemble of both spectrograms and models demonstrates significant improvement. Our best-performing system (A2, A4, A6, A7) achieves an ERR score and AuC of 0.03 and 0.994 respectively, placing in the top-3 in terms of EER score in the ASVspoof 2019 challenge \cite{top3}. These results highlight the strength of ensemble technique with leveraging multiple spectrogram analyses for feature extraction and deep learning models for pattern recognition.

\section{Conclusion}
This paper has evaluated the efficacy of a wide range of spectrograms and deep learning approaches for deepfake audio detection. By estabishling the ensemble of selective spectrograms and models, our best system achieves the EER score of 0.03 on LA dataset of ASVspoofing 2019 challenge, which is very competitive to state-of-the-art systems. Additionally, our comprehensive evaluation also indicate the potential of certain types of spectrogram (e.g. STFT) and deep learning approaches (e.g. CNN-based, finetuning pre-trained models), which can provide initial guidance for deepfake audio detection. 

%\section*{ACKNOWLEDGMENTS}
%The research leading to this publication was partially carried out within the EUCINF project. The EUCINF project is funded by the TODO program under grant no. TODO.

%\newpage
%Demo
%\begin{figure}[t]
%    	\vspace{-0.2cm}
%    \centering
%    \includegraphics[width =1.0\linewidth]{f2.png}
%	\caption{A demo of landslide segmentation on Huggingface}
%   	%\vspace{-0.4cm}
%    \label{fig:res_f2}
%\end{figure}

\addtolength{\textheight}{-1cm}   % This command serves to balance the column lengths

%\begin{thebibliography}{99}
\bibliographystyle{IEEEbib}
\bibliography{refs}
%\end{thebibliography}
\end{document}